\documentclass{aa}
\usepackage{natbib}
\usepackage{graphicx}
\bibpunct{(}{)}{;}{a}{}{,}
\begin{document}
   \title{Proper initial conditions for long-term integrations \\ of the solar system}

   \author{M. Arminjon
		  \inst{1}}

   \institute{Laboratoire "Sols, Solides, Structures"\\
CNRS / Universit\'{e} Joseph Fourier / Institut National Polytechnique de Grenoble\\
B.P. 53, 38041 Grenoble cedex 9, France. \email{arminjon@hmg.inpg.fr}}

   \date{Received date / accepted date}

   \abstract{
   An optimization program is used to re-adjust the initial conditions, in order to reproduce as closely
as possible the predictions of a complete ephemeris by using simplified equations of motion
in the numerical integration. The adjustment of the initial conditions is illustrated in the transition from
the DE406 complete long-range ephemeris to a Newtonian model considering only the Sun and the four
major planets. It is also used to best reproduce this same DE406 ephemeris, based on post-Newtonian
equations for a system of mass points and including the Moon and asteroids, by using a Newtonian calculation
corrected by the Schwarzschild effects of the Sun and restricted to the ten major bodies of the solar system.

   \keywords{Celestial mechanics -- Ephemerides}
   }

   \maketitle
%

\section{Introduction}

When integrating the equations of motion in classical or relativistic celestial mechanics, one has to know the values of the parameters, in particular the masses and the initial conditions. The masses of the celestial bodies are obtained by combining different methods, including the analysis of the perturbations that they cause to the motion of other bodies, and the fitting of the observed trajectory of a spacecraft in close approach. In the construction of modern ephemerides, the initial conditions are obtained by least-squares fittings to large sets of observational data \citep{new, fie}. It is important to realize that the optimal parameters, that lead to the minimum (least-squares) residual with respect to a set of observational data, depend on the precise model that is used. Thus, in the case that an analytical ephemeris is adjusted to a numerically-integrated one, the initial conditions taken from the numerical ephemeris have to be modified \citep{les, moi}. If one aims at testing an alternative theory of gravity, one may expect that even the masses will have to be very slightly modified \citep{arm}. A program for the adjustment of the masses and the initial conditions has been built by the author in the latter context. It has been tested by adjusting a purely Newtonian calculation to the numerically-integrated ephemeris DE403 \citep{sta95}, that is based on general-relativistic equations of motion. The program is based on the Gauss algorithm for iterative minimization of the mean quadratic error, which needs to calculate the partial derivatives of the theoretical predictions with respect to the parameters. Several standard methods may be used to compute these derivatives. In this program, they are calculated by finite differences: this involves making loops on the numerical solution of the equations of motion.

In the present paper, this optimization program is used to re-adjust the initial conditions in the transition from the complete "long" ephemeris DE406 \citep{sta98} to simplified models that are more tractable for integrations over very long times. Thus, the input data for the adjustment are taken from the complete ephemeris, and two simplified models of the solar system will be considered: in Section 2, we present the adjustment of a model based on purely Newtonian equations, and limited to the Sun and the four major planets. In Section 3, the initial conditions are re-adjusted for a model that considers the nine major planets and accounts, moreover, for those general-relativistic effects that are caused by the Sun alone. For these two adjustments, the sets of the optimal initial conditions are provided for potential users, and the differences with the predictions of the reference ephemeris are illustrated. Moreover, the effect of extrapolating the calculation outside the fitting interval can be seen, because the fitting intervals for the different planets are all significantly smaller than the 60 centuries of the DE406 ephemeris.


\section{Initial conditions for Newtonian integrations with the four major planets}

The initial conditions, at Julian Day 2\,451\,600.5 (26 February 2000, 0H00), of a Newtonian calculation considering five mass points: the Sun and the four giant planets, were adjusted. The target time interval for the adjustment was from -1000 to +2000. The input data (41 heliocentric position vectors for each planet, equally distributed in time for a given planet) were taken from the DE406 ephemeris, and spanned a good part of this interval: 1073 years for Jupiter, 2672 for Saturn, and 2913 years for Uranus and Neptune. The initial conditions were initialized at the DE406 values, then adjusted, whereas the $GM$ products were fixed at the DE406 values. The standard deviation, between the input data of DE406 and the corresponding values obtained with the Newtonian calculation restricted to the Sun and the four major planets, was reduced by a factor of 434 by the adjustment program. This ratio demonstrates the necessity of readjusting the initial conditions, as was done. Table~\ref{CI_geantes} shows the values found for the initial conditions after this adjustment.
\begin{table*} [b]
	  \caption[]{Initial conditions at JD 2\,451\,600.5 found for a Newtonian model containing only the Sun and the 4 giant planets after the adjustment  (heliocentric equatorial Cartesian coordinates).}
		 \label{CI_geantes}
\begin{tabular}{c c c c}
			\hline
Positions (au) & & & \\
Jupiter  & +3.733061134680E+00 & +3.052435941481E+00 & +1.217433227240E+00 \\
Saturn &  +6.164409988928E+00 & +6.366764207406E+00 & +2.364527399872E+00 \\
Uranus   & +1.457963610839E+01 & -1.236887103144E+01 & -5.623600018369E+00 \\
Neptune  & +1.695475862942E+01 & -2.288718303371E+01 & -9.789935831446E+00 \\
\\
\hline
Velocities (au/d) & & & \\
Jupiter  & -5.086558851107E-03 & +5.493661581130E-03 & +2.478691066898E-03 \\
Saturn   & -4.426842485459E-03 & +3.394095269623E-03 & +1.592276937224E-03 \\
Uranus   & +2.647476590726E-03 & +2.487495861475E-03 & +1.052019949681E-03 \\
Neptune  & +2.568634013660E-03 & +1.681857805109E-03 & +6.245754662095E-04 \\
\hline
\end{tabular}
\end{table*}
\begin{figure*}
   \centering
   \includegraphics[width=17cm]{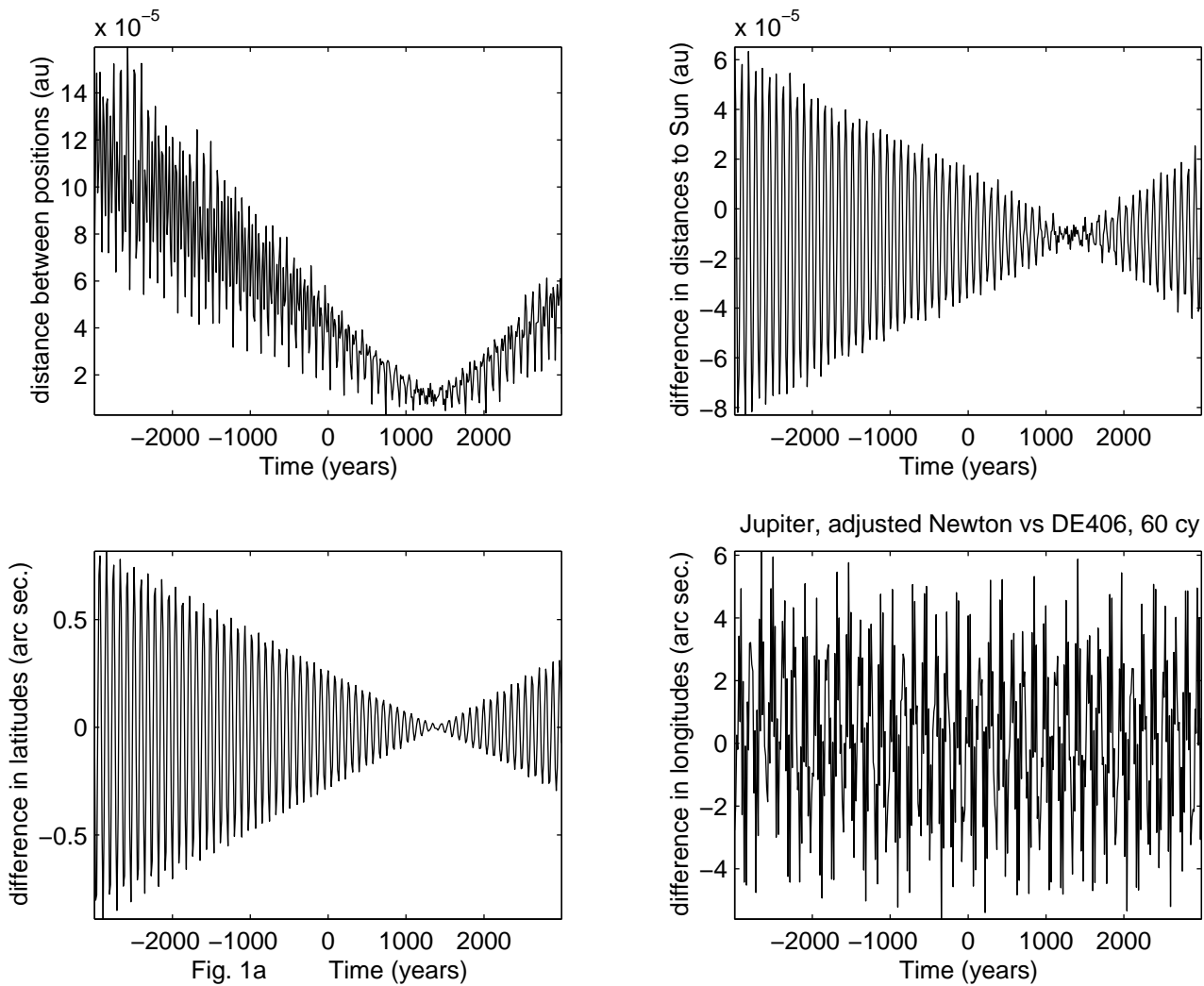}
	   \caption{ Comparison between DE406 and an adjusted Newtonian calculation restricted to the Sun and the four giant planets, over 60 centuries. Plotted are the differences between the heliocentric position vectors, distances, latitudes, and longitudes, in the sense Newtonian calculation minus DE406 ephemeris. \bf{Fig. 1a}: Jupiter.
}
			  \label{geantes}%
	\end{figure*}
\begin{figure*}
\addtocounter{figure}{-1}
   \centering
   \includegraphics[width=17cm]{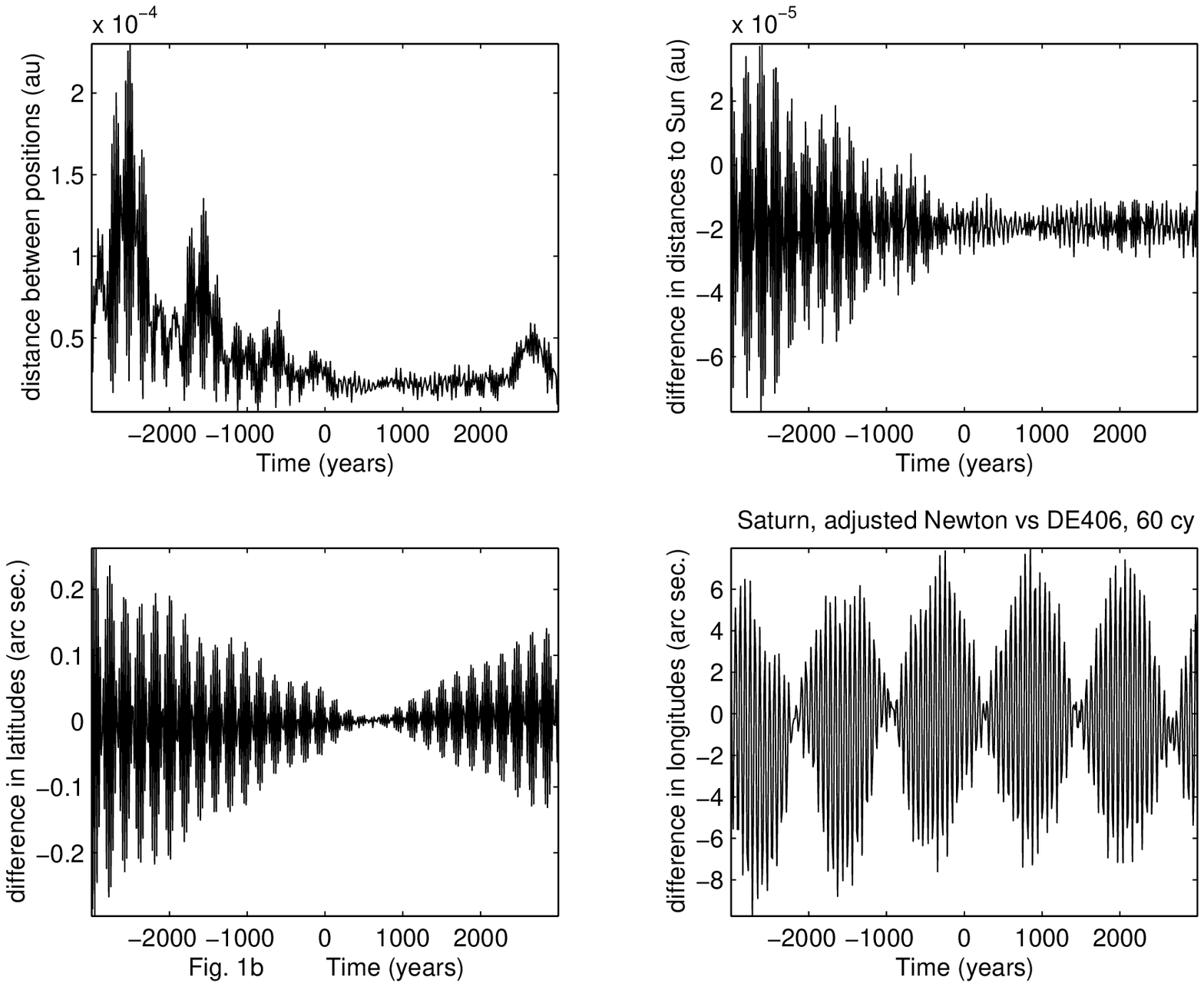}
	   \caption{ \bf{b}: Saturn.
}
	\end{figure*}
\begin{figure*}
\addtocounter{figure}{-1}
   \centering
   \includegraphics[width=17cm]{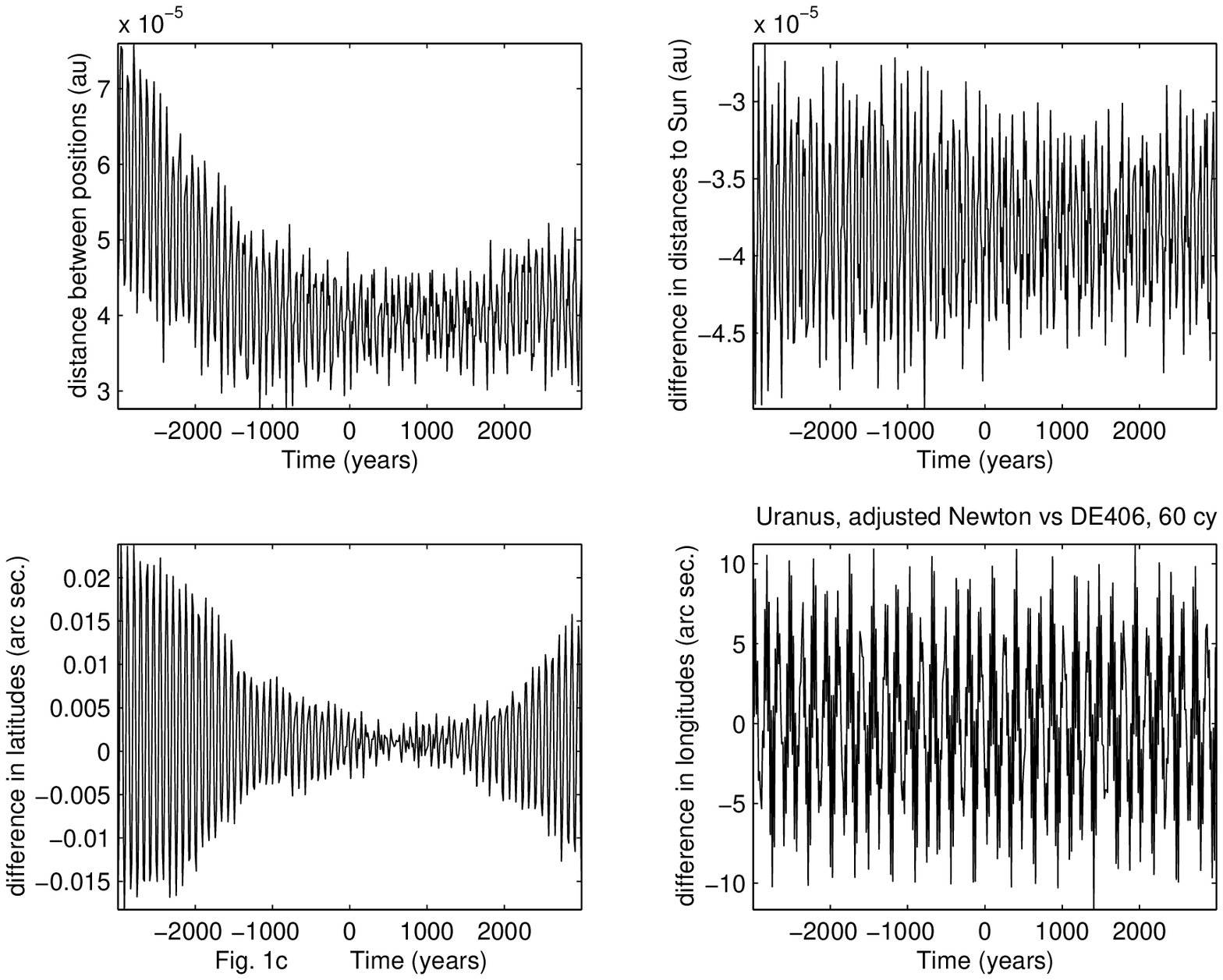}
	   \caption{ \bf{c}: Uranus.
}
	\end{figure*}
\begin{figure*}
\addtocounter{figure}{-1}
   \centering
   \includegraphics[width=17cm]{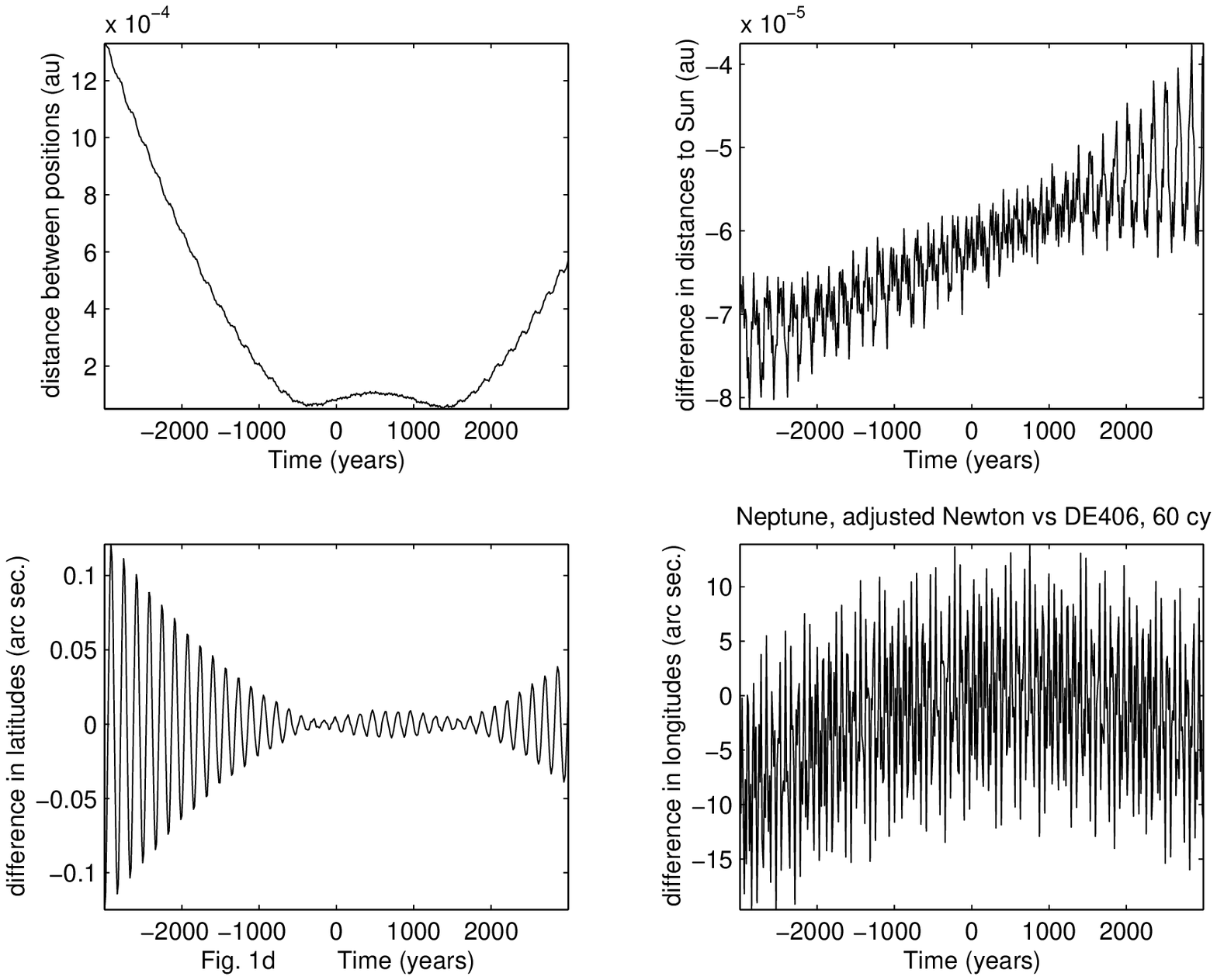}
	   \caption{ \bf{d}: Neptune.
}
			  \label{geantes}%
	\end{figure*}

Figure~\ref{geantes} shows the comparison between DE406 and this adjusted Newtonian calculation, over the 60 centuries (-3000 to +3000) covered by DE406. Heliocentric coordinates are being used for the figures and for the tables as well. The reference plane for the tables is the one which was used in the integrations, namely the Earth's J2000 equator. For the figures, the coordinate system has been rotated so that the reference plane is the invariable plane of the solar system (normal to the total angular momentum of the system). It can be seen that, using the adjusted initial conditions, the difference with DE406 remains very small, even over the extrapolated time range of 60 centuries. This is especially true if one remembers that the model used here is very strongly simplified as compared with the complete calculation of DE406. However, one observes a sensible deterioration outside the target interval of the adjustment (which, we recall, was from $-1000$ to $+2000$), especially for Neptune. The most favourable behavior in the extrapolation is found for Jupiter, for which the errors in distance and latitude increase merely linearly with time, whereas the longitude shows hardly any long-range drift. It is, of course, not possible to predict how the differences would evolve outside the 60 centuries of DE406.

\section{Initial conditions for calculations with the nine major planets, accounting for relativistic effects
 in the Schwarzschild field of the Sun}
The ephemerides series DExxx, constructed at the Jet Propulsion Laboratory, are based on Einstein-Infeld-Hoffmann-type
equations of motion, and thus include general-relativistic effects at the first post-Newtonian approximation
 in the solar system, the latter being considered as a general system of point masses \citep{new}.
If one wishes to account merely for the major general-relativistic effects, one may consider instead the Newtonian
$N$--body problem for point masses, and add, to the Newtonian acceleration of any given planet, that correction which
is obtained when considering this planet as a test particle in the Schwarzschild field of the Sun [\emph{cf.} \citet{bru}],
 thus writing the heliocentric acceleration of a planet as:
\begin{eqnarray}
\frac{\mathrm{d}\mathbf{u}}{\mathrm{d}t} & = & \textrm{heliocentric acceleration in the Newtonian} \nonumber \\
& & N\textrm{--body problem } + \nonumber \\
& & + \frac{GM}{r^{2}}\left( \left[ \frac{2}{c^{2}}\left(
\frac{GM}{r}+\frac{3}{2}\left(\mathbf{u.n}\right)^{2}-\mathbf{u}^{2}
\right)\right]\mathbf{n}\right. \nonumber \\ & & +\frac{2}{c^{2}}\left(\mathbf{u.n}\right)\mathbf{u}
\left. \right)
\label{accelPN}
\end{eqnarray}

where $G$ is the gravitation constant and $c$ is the velocity of light, $M$ is the mass of the Sun, $r\equiv\|\mathbf{x}\|\equiv(\mathbf{x.x})^{1/2}$, $\mathbf{n}\equiv\mathbf{x}/\|\mathbf{x}\|$,
and $\mathbf{u}\equiv\mathrm{d}\mathbf{x}/\mathrm{d}t$, $\mathbf{x}$ being the heliocentric position vector.
(Equation~(\ref{accelPN}) applies when Schwarzschild's solution is expressed in standard coordinates.) \citet{les} proposed an analytical integration of Eq.~(\ref{accelPN}), based on a Taylor expansion of Gauss' perturbation equations for the osculating elements. They presented numerical results for Mercury, and found very small differences with the DE102 ephemeris, after the integration constants of their integration had been adjusted on DE102.
In the present work, a numerical integration of Eq.~(\ref{accelPN}) has been used, based on the $\mathrm{Matlab}$ solver ODE113. [For all calculations in this paper, the same numerical tolerances have been used in the ODE113 routine as those found optimal in our first test of this program, {\em i.e.} $\mathrm{RelTol}=5\times10^{-13}$ and $\mathrm{AbsTol}=10^{-15}$.
Moreover, using the osculating elements, some analytical perturbations of the Moon on the motion of the EMB \citep{bre} were taken into account in the program.] We found only a marginal increase in the computation time when passing from the Newtonian heliocentric equations of motion to Eq.~(\ref{accelPN}). The reasons are that: i) the program spends the greatest part of the time in calculating the Newtonian accelerations due to the other planets, and ii) when the "Schwarzschild corrections" are added, the set of the step sizes automatically adopted by the solver in order to reach the given tolerances, does not have a smaller average than has the set of the step sizes adopted with purely Newtonian equations (and with the same tolerances). The latter point means that the Schwarzschild corrections do not "stiffen" the differential system.

However, the integration time is roughly 200 times greater with ten bodies including Mercury, than for five bodies with Jupiter as the quickest planet (the latter case was investigated in Section 2), in addition there are more initial conditions to adjust. For this reason, only "short-range" adjustments may be run easily with common computers. For the present adjustment, 21 positions per planet were taken from the DE406 ephemeris. As for the previous adjustment, the time intervals spanned by these input data increased with the period of the planet: 4000 days for Mercury, and 64000 days (175 years) for the five outermost planets. The adjustment program reduced the standard deviation between the input data of DE406 and the corresponding values obtained with the calculation based on Eq.~(\ref{accelPN}), by a factor of 492. This again demonstrates the necessity of readjusting the initial conditions. Table~\ref{CI_toutes} gives the initial conditions found after this adjustment.

\begin{table*} [b]
	  \caption[]{Initial conditions at JD 2\,451\,600.5 found for the model based on Eq.~(\ref{accelPN}).}
		 \label{CI_toutes}
\begin{tabular}{c c c c}
			\hline
Positions (au) & & &\\
Mercury & -2.503321047836E-01 & +1.873217481656E-01 & +1.260230112145E-01  \\
Venus & +1.747780055994E-02 & -6.624210296743E-01 & -2.991203277122E-01 \\
Earth-Moon Baryc. & -9.091916173950E-01 & +3.592925969244E-01 & +1.557729610506E-01 \\
Mars & +1.203018828754E+00 & +7.270712989688E-01 & +3.009561427569E-01 \\
Jupiter & +3.733076999471E+00 & +3.052424824299E+00 & +1.217426663570E+00 \\
Saturn & +6.164433062913E+00 & +6.366775402981E+00 & +2.364531109847E+00 \\
Uranus & +1.457964661868E+01 & -1.236891078519E+01 & -5.623617280033E+00 \\
Neptune & +1.695491139909E+01 & -2.288713988623E+01 & -9.789921035251E+00 \\
Pluto & -9.707098450131E+00 & -2.804098175319E+01 & -5.823808919246E+00 \\
\\
\hline
Velocities (au/d) & & & \\
Mercury & -2.438808424736E-02 & -1.850224608274E-02 & -7.353811537540E-03 \\
Venus & +2.008547034175E-02 & +8.365454832702E-04 & -8.947888514893E-04 \\
Earth-Moon Baryc. & -7.085843239142E-03 & -1.455634327653E-02 & -6.310912842359E-03 \\
Mars & -7.124453943885E-03 & +1.166307407692E-02 & +5.542098698449E-03 \\
Jupiter & -5.086540617947E-03 & +5.493643783389E-03 & +2.478685100749E-03 \\
Saturn & -4.426823593779E-03 & +3.394060157503E-03 & +1.592261423092E-03 \\
Uranus & +2.647505630327E-03 & +2.487457379099E-03 & +1.052000252243E-03 \\
Neptune & +2.568651772461E-03 & +1.681832388267E-03 & +6.245613982833E-04 \\
Pluto & +3.034112963576E-03 & -1.111317562971E-03 & -1.261841468083E-03 \\
\hline
\end{tabular}
\end{table*}
\begin{figure*}
   \centering
   \includegraphics[width=17 cm]{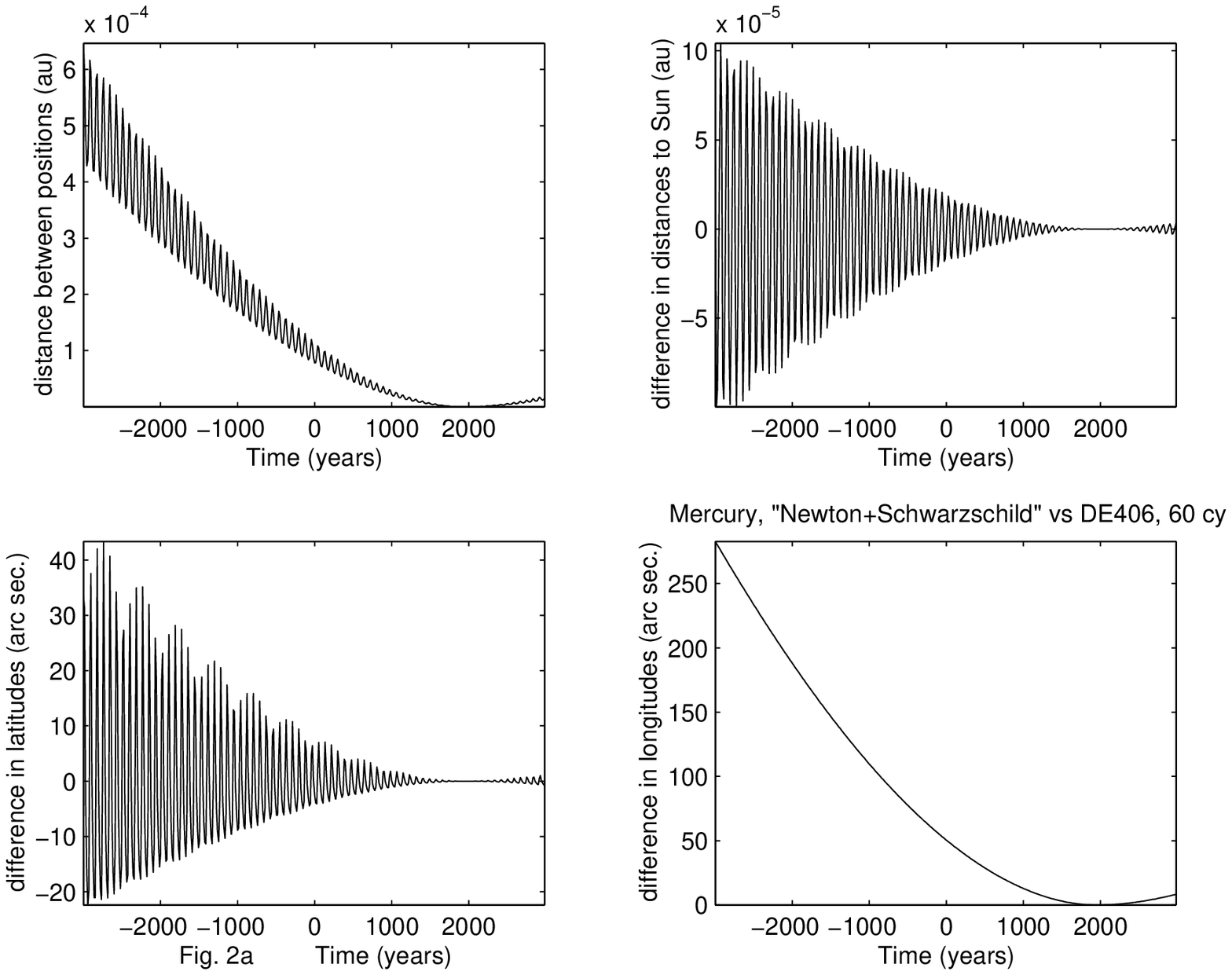}
	   \caption{ Comparison between DE406 and an adjusted calculation restricted to the Sun and the nine major planets,
and including Schwarzschild corrections of the Sun, over 60 centuries. Plotted are the differences between the heliocentric
 position vectors, distances, latitudes, and longitudes, in the sense (Newtonian calculation with Schwarzschild corrections)
 minus (DE406 ephemeris). \bf{Fig. 2a}: Mercury.
}
			  \label{toutes}%
	\end{figure*}
\begin{figure*}
\addtocounter{figure}{-1}
   \centering
   \includegraphics[width=17 cm]{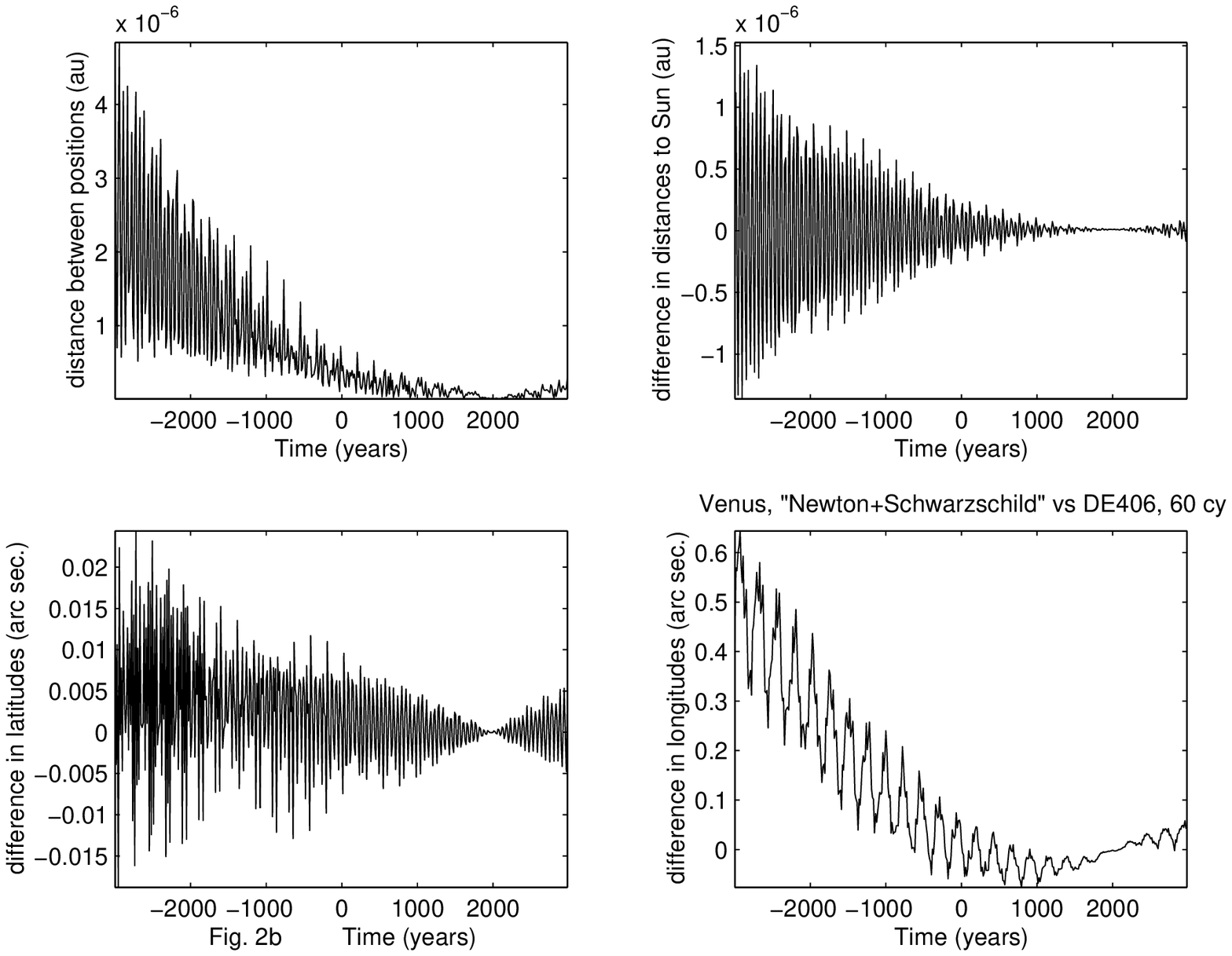}
	   \caption{ \bf{b}: Venus.
}
	\end{figure*}
\begin{figure*}
\addtocounter{figure}{-1}
   \centering
   \includegraphics[width=17 cm]{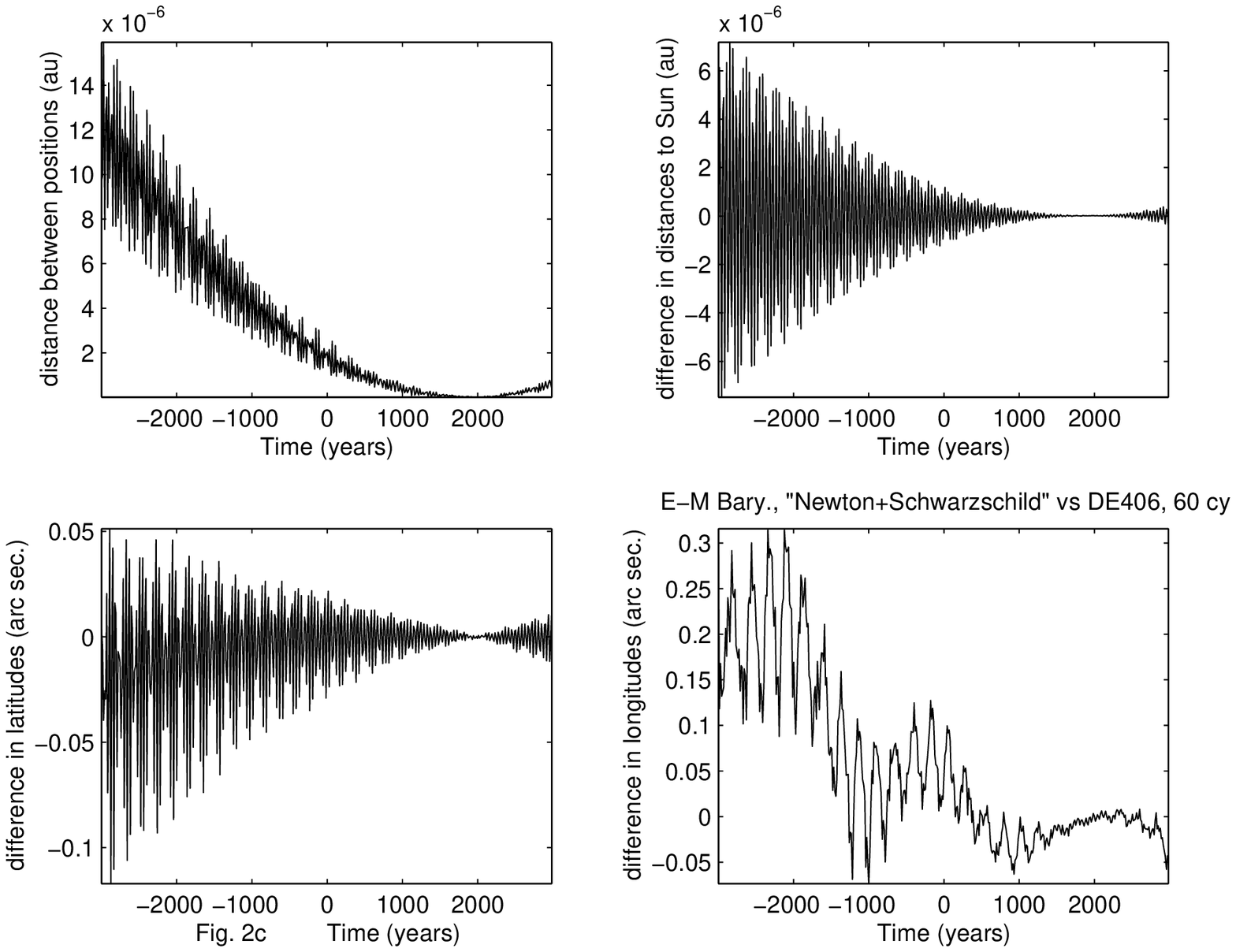}
	   \caption{\bf{c}: Earth-Moon Barycenter.
}
	\end{figure*}
\begin{figure*}
\addtocounter{figure}{-1}
   \centering
   \includegraphics[width=17 cm]{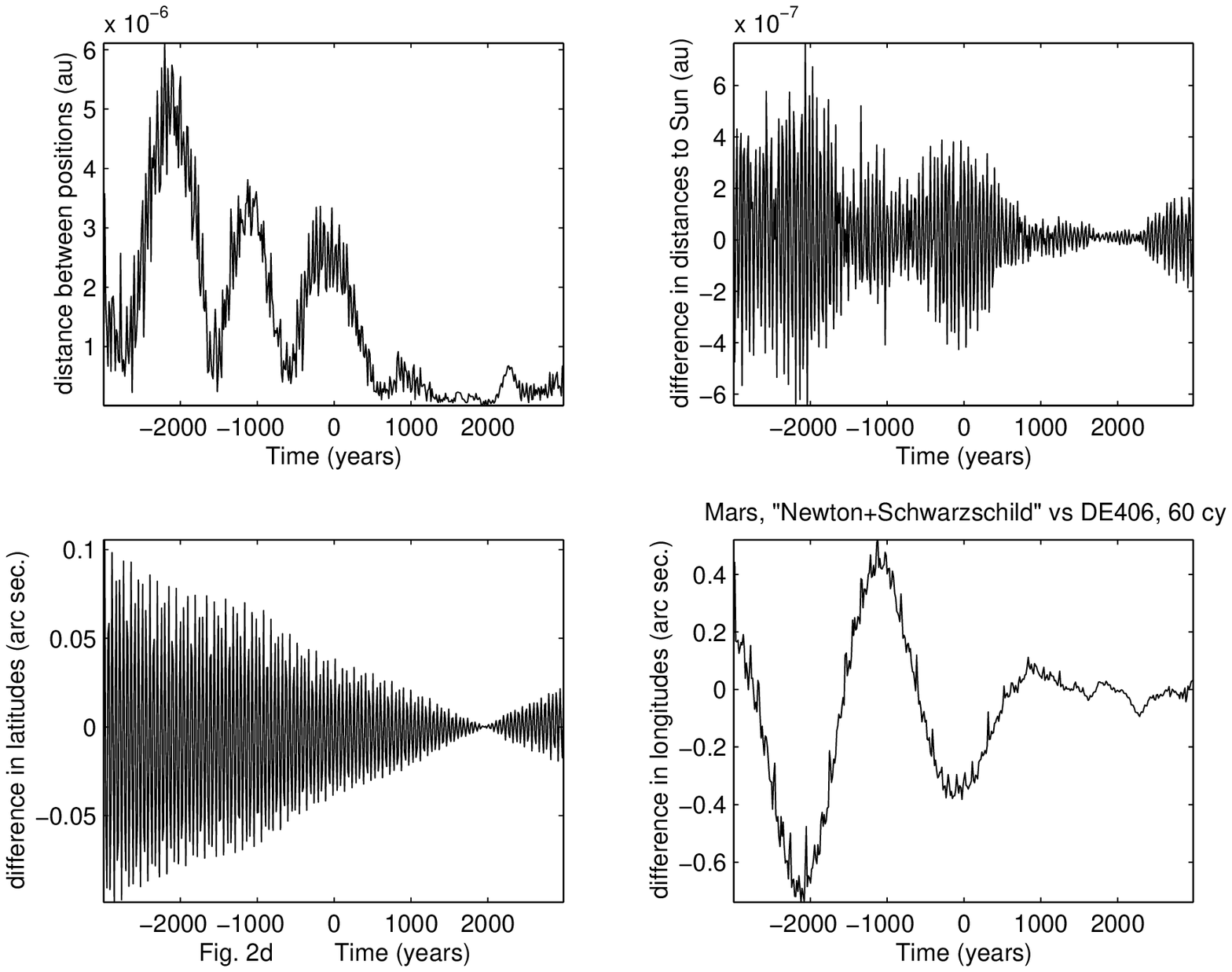}
	   \caption{\bf{d}: Mars.
}
	\end{figure*}
\begin{figure*}
\addtocounter{figure}{-1}
   \centering
   \includegraphics[width=17 cm]{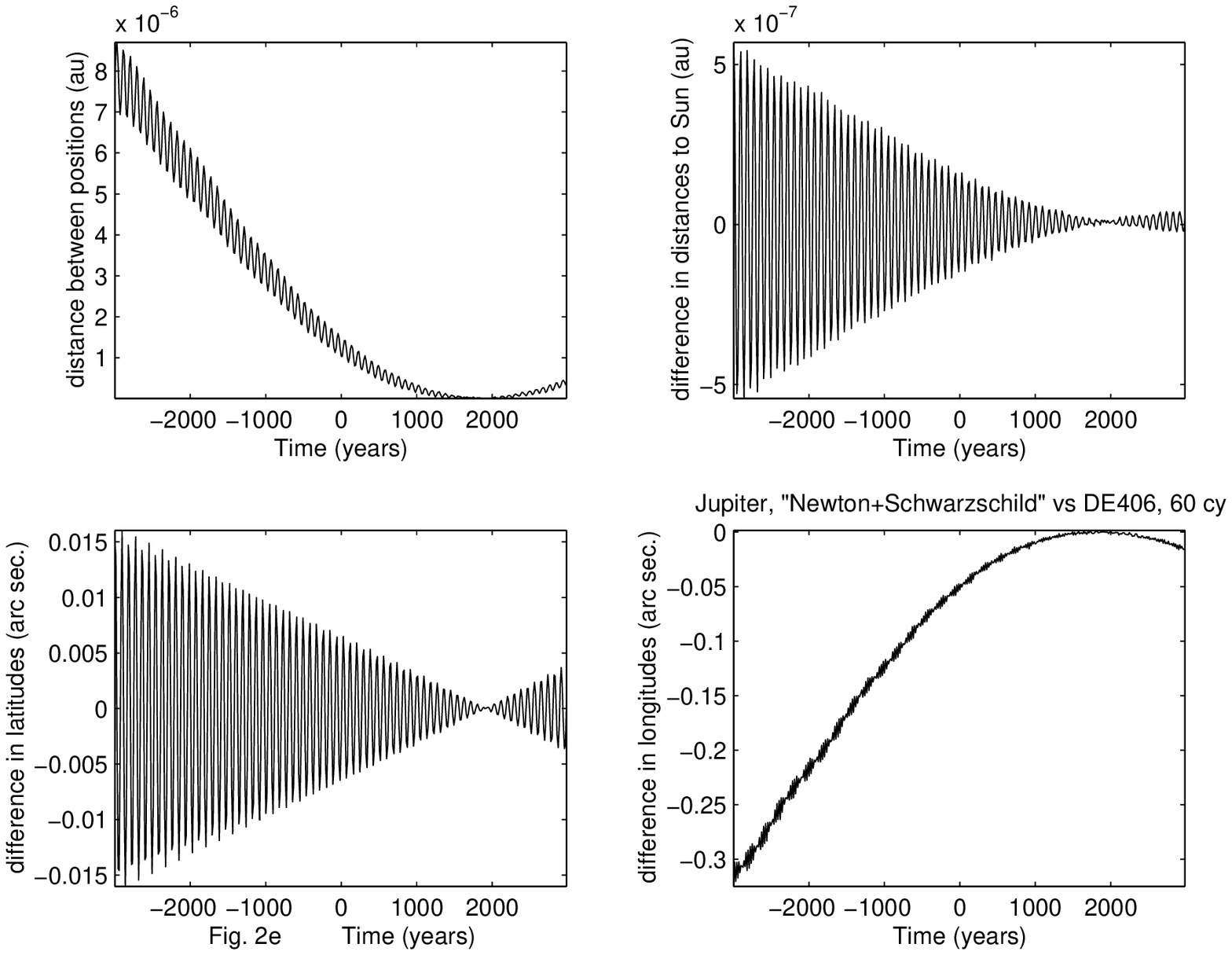}
	   \caption{\bf{e}: Jupiter.
}
	\end{figure*}
\begin{figure*}
\addtocounter{figure}{-1}
   \centering
   \includegraphics[width=17 cm]{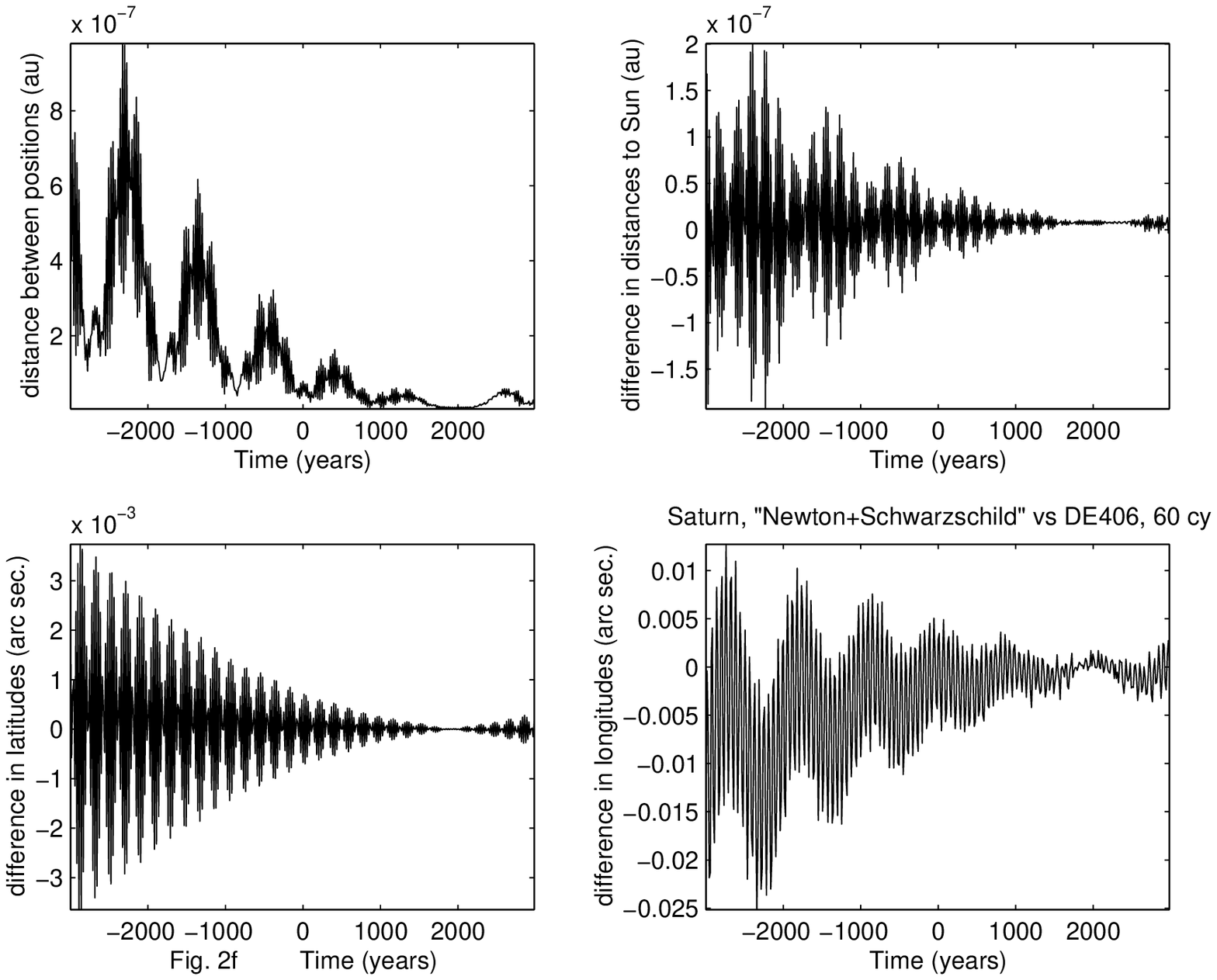}
	   \caption{ \bf{f}: Saturn.
}
	\end{figure*}
\begin{figure*}
\addtocounter{figure}{-1}
   \centering
   \includegraphics[width=17 cm]{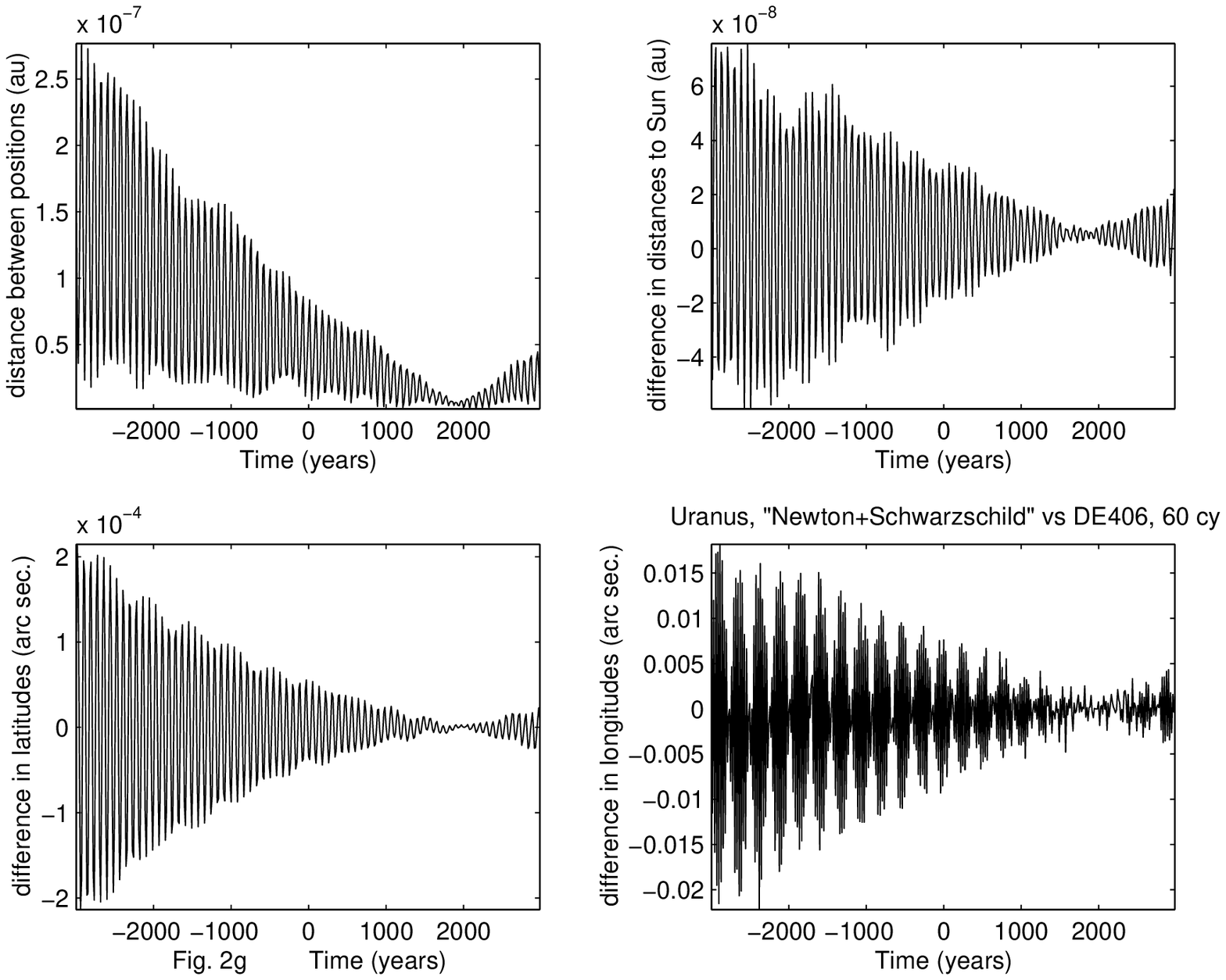}
	   \caption{ \bf{g}: Uranus.
}
	\end{figure*}
\begin{figure*}
\addtocounter{figure}{-1}
   \centering
   \includegraphics[width=17 cm]{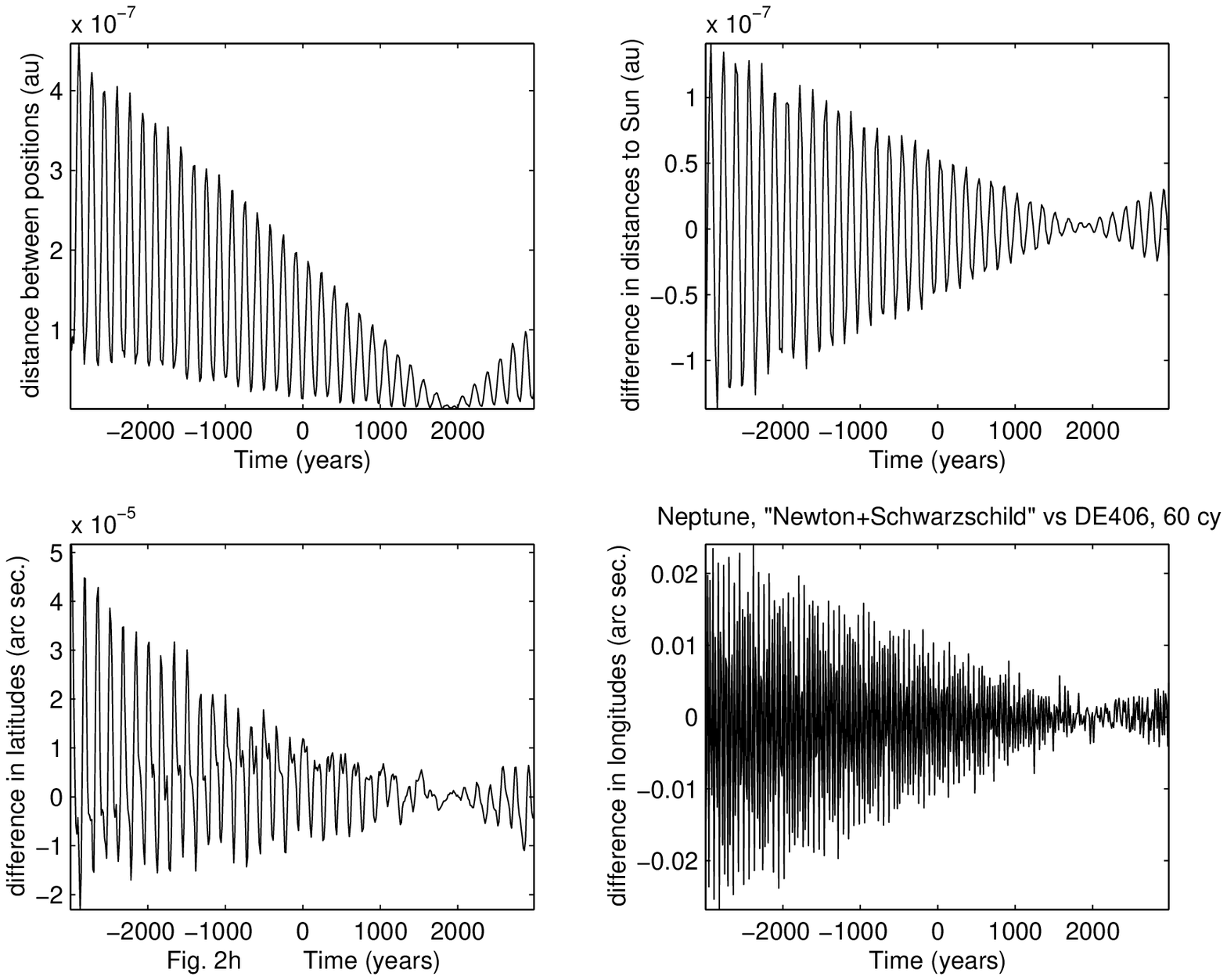}
	   \caption{\bf{h}: Neptune.
}
	\end{figure*}
\begin{figure*}
\addtocounter{figure}{-1}
   \centering
   \includegraphics[width=17 cm]{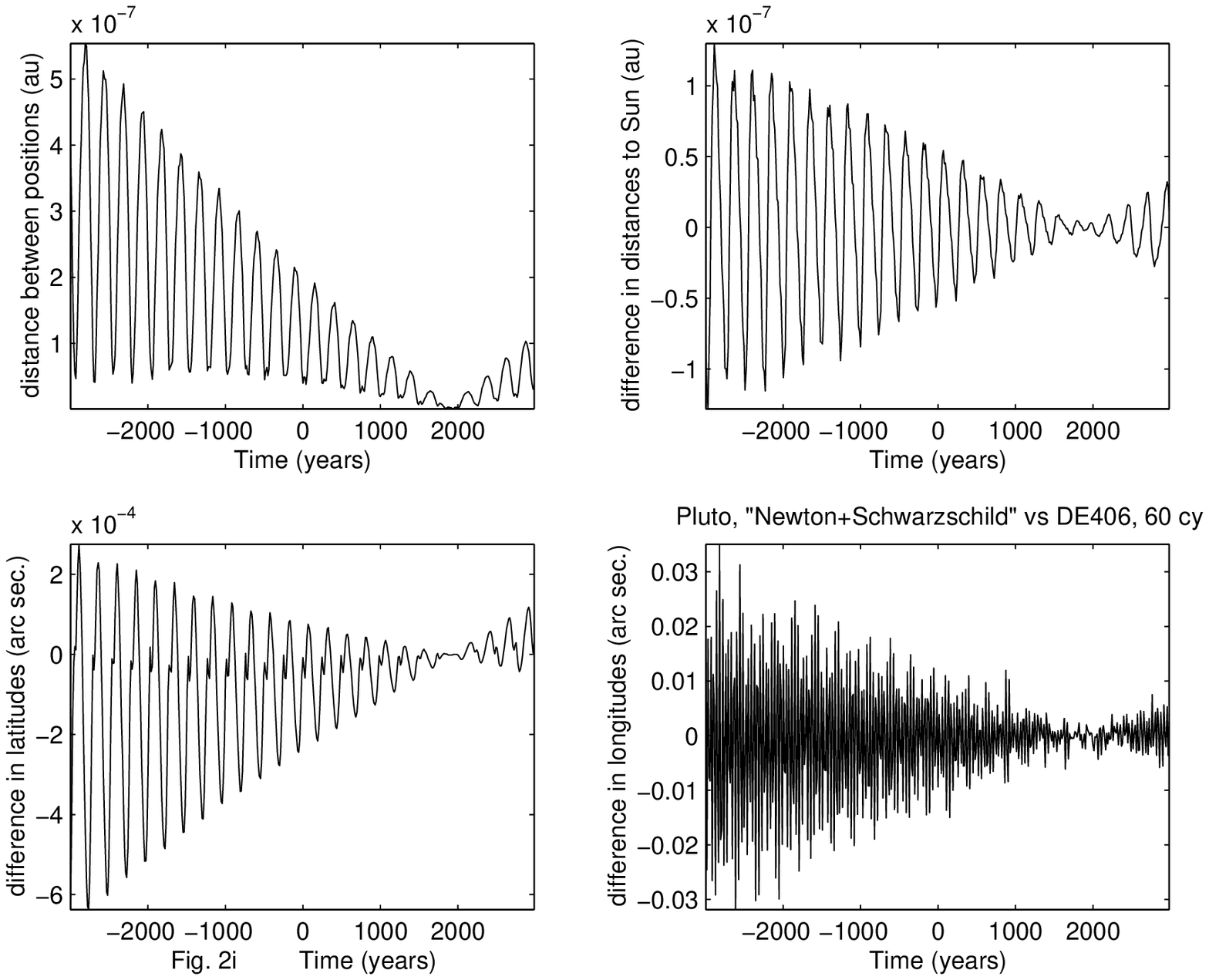}
	   \caption{\bf{i}: Pluto.
}
	\end{figure*}
Figure~\ref{toutes} shows the comparison between DE406 and this adjusted Newtonian calculation with Schwarzschild corrections, over the 60 centuries of DE406. If one first compares Figures 2e-f-g-h with Figures 1a-b-c-d, one observes that, for the giant planets, the present calculation gives gains of accuracy (the latter being measured, e.g., by the norm $\delta R$ of the radius vector difference) by a factor of: 20, 250, 300, and 3000, for Jupiter, Saturn, Uranus, and Neptune, respectively, as compared with the simpler calculation of Section 2. The poorest accuracy is found for Mercury (Fig. 2a). Over one century, the error is very small for Mercury also: $6\times10^{-8}$ au for $\delta R$ and $0.03\arcsec$ for the longitude, but it increases more quickly with time for Mercury. This may be due, at least for one part, to the numerical integration error.

The effect of changing the weights was considered: by taking weights that are proportional to the inverse of the data uncertainty, one reduces the standard deviations between input data and their recalculated values, by a factor two for Venus and the EMB, by 50\% for Mars, and by 25\% for Mercury, although the residuals for outer planets are all deteriorated (by factors of ten for Jupiter, and of several hundreds for Uranus, Neptune and Pluto). In order to still improve the agreement with a complete ephemeris, one could also consider other fine points, e.g. whether the same form of metric (isotropic or harmonic, etc.) is being used in the Schwarzschild solution as in the post-Newtonian equations of motion for a system of mass points. But, if the accuracy over long time ranges is aimed at, one should first check the effect of improving the accuracy of the numerical integration (e.g. by passing from double to quadruple precision).

\section{Conclusion}
The adequate values of all parameters in the equations of motion depend on the precise form adopted for the latter equations. In particular, it is very important to re-adjust the initial conditions when one passes from a complete form of the equations of motion, as it is used to build reference ephemerides, to a simplified form, which may be more adequate for long-range calculations. Although it may be too time-consuming to optimize the initial conditions so as to get the smallest residual for the whole time-range to be considered, one can get "reasonably extrapolable" initial conditions, by optimizing them for a time interval that contains at least a few periods of even the slowest body considered. Provided one thus re-adjusts the initial conditions, it is possible to reproduce quite accurately the corrections of post-Newtonian general relativity, by adding the Schwarzschild corrections, that account for the post-Newtonian effects of the Sun alone. This does not significantly increase the computer time as compared with the purely Newtonian equations of motion. A very precise comparison between the complete and simplified ways of describing post-Newtonian effects in the solar system could be obtained only if the two models were identical in the other respects, which was not the case here. Including the masses in the solved-for parameters brought only marginal improvements in the models for which it was considered.


\begin{acknowledgements}
	 This work was suggested by E. M. Standish. I am grateful to him for his careful reading of the paper and for his help in the discussion of the results. I would also like to thank P. Bretagnon and C. Le Guyader for their help at an earlier stage of this research.
\end{acknowledgements}

\bibliographystyle{aa}

\end{document}